\documentstyle[12pt,aasms4]{article}

\begin{document}

\title{EGRET Spectral Index and the Low-Energy Peak Position in the
        Spectral Energy Distribution of EGRET-Detected Blazars}

\author{Y.~C.~Lin\altaffilmark{1,2}, D.~L.~Bertsch\altaffilmark{3},
 S.~D.~Bloom\altaffilmark{4}, J.~A.~Esposito\altaffilmark{5}, 
 R.~C.~Hartman\altaffilmark{3}, S.~D.~Hunter\altaffilmark{3}, 
 G.~Kanbach\altaffilmark{6}, D.~A.~Kniffen\altaffilmark{7}, 
 H.~A.~Mayer-Hasselwander\altaffilmark{6}, P.~F.~Michelson\altaffilmark{1}, 
 R.~Mukherjee\altaffilmark{8}, A.~M.~M\"ucke\altaffilmark{9}, 
 P.~L.~Nolan\altaffilmark{1}, M.~Pohl\altaffilmark{10}, 
 O.~Reimer\altaffilmark{6}, E.~J.~Schneid\altaffilmark{11}, 
 D.~J.~Thompson\altaffilmark{3}, W.~F.~Tompkins\altaffilmark{1}}

\altaffiltext{1}{W.~W.~Hansen Experimental Physics Laboratory,
 Stanford University, Stanford CA 94305 USA}
\altaffiltext{2}{lin@egret0.stanford.edu}
\altaffiltext{3}{Code 661, Laboratory for High Energy Astrophysics, 
 NASA Goddard Space Flight Center, Greenbelt, MD 20771 USA}
\altaffiltext{4}{IPAC, Jet Propulsion Laboratory, California Institute of 
 Technology, Pasadena, CA 91125 USA}
\altaffiltext{5}{Research and Data Systems Corporation, 7501 Forbes Blvd.,
 Suite 104, Seabrook, MD 20706 USA}
\altaffiltext{6}{Max-Planck-Institut f\"ur Extraterrestrische Physik,
Giessenbachstr D-85748 Garching Germany}
\altaffiltext{7}{Department of Physics, Hampden-Sydney College,
 Hampden-Sydney, VA 23943 USA}
\altaffiltext{8}{Barnard College and Columbia University, 
 New York, NY 10027 USA}
\altaffiltext{9}{Dept.~of Phys.~and Mathematical Phys., Univ.~of 
Adelaide, Adelaide, SA 5005 Australia}
\altaffiltext{10}{Institut f\"ur Theoretische Physik 4, Ruhr-Universit\"at
Bochum, 44780 Bochum Germany}
\altaffiltext{11}{Northrop-Grumman Aerospace Corporation, Mail Stop A01-26, 
 Bethpage, NY 11714 USA}

\begin{abstract}       
In current theoretical models of the blazar subclass of active
galaxies, the broadband emission consists of two components: a
low-frequency synchrotron component with a peak in the IR to
X-ray band, and a high-frequency inverse Compton component with a
peak in the gamma-ray band.  In such models, the gamma-ray
spectral index should be correlated with the location of the
low-energy peak, with flatter gamma-ray spectra expected for
blazars with synchrotron peaks at higher photon energies
and vice versa.  Using the EGRET-detected blazars as a sample, 
we examine this correlation and possible uncertainties in its 
construction.
\end{abstract}


\section{Introduction}

It is now generally believed in a class of theoretical models 
(see e.g.~Ulrich, Maraschi, \& Urry~1997 for a review)
that the broadband spectrum of a blazar consists of two distinctive
components: (a) a low-energy component which is the result of 
synchrotron radiation of a beam of relativistic particles, and which 
peaks, in the spectral energy distribution (SED) plot, in the IR to soft 
X-ray region; (b) a high-energy component which is the result of 
inverse Compton scattering of the same beam of relativistic particles 
on some ambient field of soft photons, and which peaks in SED in the
MeV-GeV-TeV region.  These models are well known for their attempt to
explain the most salient features of the broadband spectra of blazars 
from radio energies all the way to the TeV energies, an energy span 
of more than 20 orders of magnitude.

In this paper we examine an important prediction of this class of
theoretical models with EGRET data.  The two broad peaks in SED of a
blazar as described in such models, being the products of the same beam
of relativistic particles, should be closely related to each other.  
Since the high-energy peaks in SED of various blazars pass through 
the EGRET energy range from $\sim$~30~MeV to 20~GeV, the spectral 
shapes of the EGRET-detected blazars in the EGRET energy range should 
change systematically with respect to the positions of the low-energy 
SED peaks in different objects.  This prediction of these currently 
investigated theoretical models can be tested with EGRET data.

A brief and preliminary result of a study of this kind with EGRET data,
based on the Second EGRET Catalog and its Supplement 
(Thompson et al.~1995, Thompson et al.~1996), has already been sent for
publication in the Proceedings of the BL~Lac Phenomenon Meeting of 1998 
in Turku, Finland (Lin et al.~1998).  In the present paper, we expand 
the scope of the previous study with additional information taken from 
the recently published Third EGRET Catalog (Hartman et al.~1999) 
and other publications to examine again in more details the question 
of the possible correlation between EGRET spectral shapes and low-energy 
SED peak positions for the blazars that have been detected by EGRET. 

\section{The Data}

We select 27 EGRET-detected blazars (Fichtel et al.~1994, 
Thompson et al.~1995, Thompson et al.~1996, Hartman et al.~1999).
These are the ones for which the SED can be found in the published
data at least to the extent that the low-energy peak positions can be 
determined, and for which the EGRET photon spectra can also be 
determined.  Four of the sources in this sample are traditionally 
regarded as X-ray-selected BL~Lac objects 
(XBL, see Ciliegi et al.~1995).  Recently these objects have been 
reclassified as high-energy peaked BL~Lac objects 
(HBL, see Ulrich, Maraschi, \& Urry~1997).  Another eleven of these 
sources are usually regarded as radio-selected BL Lac objects 
(RBL, see Ciliegi et al.~1995), or reclassified as low-energy peaked 
BL Lac objects (LBL, see e.g.~Ulrich, Maraschi, \& Urry~1997).  
The other twelve sources in the sample belong to what are generally 
referred to as flat-spectrum radio quasars (FSRQ).  For two of the 
four XBL (HBL), Mrk~501 and PKS~2005$-$489, the EGRET detections are 
somewhat weak but still fairly certain (Kataoka et al.~1999,
Sreekumar et al.~1999, Lin et al.~1997).  Three of the four XBL have 
been detected at TeV energies (see e.g. Krennrich et al.~1999 or 
Macomb et al.~1995 for Mrk~421, Kataoka et al.~1999 or 
Kennrich et al.~1999 for Mrk~501, and Chadwick et al.~1999 for 
PKS~2155$-$304), while a good TeV flux upper limit exists for the 
fourth one (PKS~2005$-$489, Roberts et al.~1998).  Thus the SED of 
these four XBL (HBL) can be constructed well into the TeV energies 
with the high-energy peaks clearly seen.  Furthermore, five of the 
EGRET-detected FSRQ have also been detected by OSSE and COMPTEL in 
the 0.05 to 15~MeV energy range.  Combined spectra have been determined 
for these five sources through the OSSE/COMPTEL/EGRET energy ranges
(3C~273, 3C~279, CTA~102, PKS~0528+134, and 3C~454.3, see 
McNaron-Brown et al.~1995).  These five sources are all included in the 
sample here.  The high-energy peaks of these five sources can be 
constructed in the MeV to GeV energy range with strict simultaneous data 
(McNaron-Brown et al.~1995).  These peak positions are visible 
as spectral break points between the OSSE/COMPTEL data and the EGRET data.  

This sample of 27 EGRET-detected blazars are listed in Table~1 
together with EGRET fluxes, EGRET spectral indices, and the 
low-energy SED peak frequencies.  The EGRET fluxes and spectral 
indices are taken from the recent Third EGRET Catalog 
(Hartman et al.~1999) unless noted otherwise.  Most of the sources in 
the EGRET catalogs carry multiple flux values.  The values quoted here 
in Table~1 are the first entries in the Third EGRET Catalog 
upon which the source positions and the source identifications are 
determined.  The EGRET spectral indices in the Third EGRET Catalog 
are those determined for the sum data of Cycle~1 through Cycle~4
(1991 April~22 to 1995 October~4).  There are some evidence that
the EGRET spectra of some blazars tend to become harder at higher
flux levels (Mukherjee et al.~1997).  But the variations are small
and only become apparent for bright EGRET sources when the spectral 
indices can be determined with high degrees of accuracy.  So the EGRET
spectral indices listed in the Third EGRET Catalog and quoted here
in Table~1, though calculated only as average values over long period
of time, are good representation of the actual spectral shapes.
The low-energy SED peak frequencies in Table~1 are taken from 
published data.  References for these information are given in the 
footnotes below the table.  On the average, the peak positions can be 
determined from the published data to an accuracy of about 
$\rm \pm 0.2$ in the scale of $\rm log_{10}(frequency)$.  Out of the
27 sources studied here, four are found to have enough data to show
the low-energy SED peak frequency at different epochs: 
PKS~0235+164, Mrk~421, 3C~279, and BL~Lacertae (see references cited in
Table~1).  The ranges of the 
peak frequencies of these four sources are 0.5 for PKS~0235+164, 
1.0 for Mrk~421, 0.3 for 3C~279, and 0.2 for BL~Lacertae, in the scale of 
$\rm log_{10}(frequency)$.  These variations in the low-energy SED 
peak frequencies are small compared with the full frequency range of all
blazars studied here.  For these four sources, we enter the peak 
frequencies corresponding to quieter times in Table~1, 
as these are the situations where more abundant data are available.
Finally some special features of the individual sources 
are included as remarks in the last column of Table~1.  This source 
sample is not meant to be a complete one.  We just try to construct a 
sample size that is sufficiently large to draw certain statistical 
conclusions.


\section{The Analysis}

To examine the correlation between the low-energy peak and the 
high-energy peak in the SED of a blazar, we should ideally try to 
match these two peaks over broad energy ranges that cover 
substantial segments of the spectrum.  But this is not feasible
at present with existing data.  Only a handful of sources have 
detailed measurements of the spectra from radio energies to 
TeV energies.  Conclusions drawn from these few sources are likely 
to be biased in some way and not generally applicable to blazars as 
a class.  Most of the blazars in the literature have good measurements 
on their broadband spectra only around the low-energy SED peaks.  
For the EGRET-detected blazars in general, there are no existing 
data to show where the high-energy peak frequencies are located 
except for the few sources that are either the XBL mentioned above 
or the ones that have also been detected by OSSE and COMPTEL 
(McNaron-Brown et al.~1995), also mentioned above.  To examine the 
correlation between the two SED peaks, we need to rely on some specific 
properties of the broadband spectra in the two energy regions.  

For the low-energy peaks in SED, it is natural to examine the peak 
frequencies as these are the prominent spectral features that can be 
determined fairly accurately from published data.  Then in the
EGRET energy range we examine the spectral shapes, or more specifically
the spectral indices to see if they systematically change with the
low-energy peak frequencies.  

In addition to examination of the model prediction on the correlation
between the low-energy SED peak frequency and the EGRET spectral index,
which involves only experimental data as described above, we can also 
compare the observed EGRET spectral indices with the calculated spectral 
indices from these theoretical models of blazars currently under 
investigation (see e.g.~Ulrich, Maraschi, \& Urry~1997) in the EGRET 
energy region.  We take the illustrative theoretical curves plotted 
in Figure~12 of Fossati et al.~(1999) and determine the slopes of these 
curves at 100~MeV as a function of the low-energy SED peak frequencies 
in the figure.  We then compare these theoretical slopes with EGRET 
spectral indices of the sources listed in Table~1 as functions of the 
low-energy peak frequencies.


In Figure~1 we plot the EGRET spectral index $\rm \gamma - 2$ 
($\rm f \sim E^{-\gamma}$) vs. the low-energy SED peak frequency in 
$\rm log_{10}$ scale for the sample of 27 EGRET-detected blazars 
examined in this paper.  The source designations, their low-energy SED 
peak frequencies, and the corresponding EGRET spectral indices are all 
listed in Table~1.  The value $\rm \gamma -2$ corresponds to the 
spectral index in an SED plot.  The five FSRQ that are also detected
by OSSE/COMPTEL are indicated separately in the figure.  In Figure~1 we 
also plot the theoretical prediction of the spectral slope at 100~MeV 
as a function of the low-energy SED peak frequency as described in
the paragraph above.  Some of the graph points in Figure~1 are slightly 
shifted in their abscissae to avoid graph congestion.  The theoretically 
calculated spectral slopes at 100~MeV are connected with dotted lines.

As one can see in Figure~1, the EGRET spectral indices do 
not vary systematically with respect to the low-energy SED peak 
frequencies.  The four XBL (HBL) may form the only exception to 
this general situation.  The theoretical models seem to work well 
for the four EGRET-detected XBL.  But when the low-energy SED peak 
frequency moves toward lower values, the agreement between the 
theoretical predition and the EGRET data ceases to exist.  The EGRET 
spectral indices do not form a pattern that may resemble the upturn 
of the theoretical curve toward smaller low-energy SED peak frequencies.  
One cannot find a trend of other kind either in the EGRET spectral 
indices in Figure~1.  The error margins in the EGRET spectral 
indices cannot accommodate the discrepancy between the theoretical 
prediction and the EGRET data.  Many of the EGRET spectral indices 
have been determined with an accuracy better than $\rm \sim \pm 0.15$.

We may also add that for the four EGRET-detected XBL (HBL), the EGRET 
spectral indices $\gamma$ ($\rm f \sim E^{-\gamma}$) all fall within 
the low range between 1.5 and 1.7 while the low-energy SED peak 
frequencies all fall within the high range between $\rm 10^{16}$
and $\rm 10^{17}$~Hz.  This fact is consistent with the results
of the theoretical model fits (see e.g. Fossati et al.~1999).  
But when it comes to the individual spectral and peak frequency values, 
there is no correlation between the EGRET spectral indices and the 
low-energy SED peak frequencies for these four XBL either.  However 
we must point out that the error margins in the EGRET spectral indices 
are large for these four XBL and the lack of correlation found here 
does not carry much weight.  But for the five EGRET-detected FSRQ that 
have also been detected by OSSE/COMPTEL, and as such the high-energy 
SED peak frequencies can also be determined with the combined 
OSSE/COMPTEL/EGRET data as the spectral break points in the 0.015~MeV 
to several GeV energy range (see Figure~2 in McNaron-Brown et al.~1995), 
these five sources do not show correlation between the low-energy SED 
peak frequencies and the high-energy SED peak frequencies.  The blazars 
that can be detected by OSSE or COMPTEL are likely to be those with 
steep EGRET spectra because then the EGRET spectra will extend
high into the COMPTEL/OSSE energy regions.  We may expect to see better
agreement between theory and data for these five sources alone where the
EGRET spectral indices become high.  But in Figure~1,
these five sources do not follow the theoretical curve either.

\section{Discussion}

Many of the EGRET-detected blazars suffer from poor statistical accuracy
because of limited photon counts; in those cases, the spectral indices
are very poorly determined.  However, in Figure~1 it is apparent
that the greatest discrepancy between the data and the theoretical
prediction is at the lowest synchrotron-peak frequencies, where most of
the best-determined EGRET indices are found.  In the unified blazar
scenarios it might be expected that the objects with synchrotron-peak
frequencies below $10^{13}$~Hz are all FSRQ's; however, note that {\it
all\/} of these are {\it far\/} below the (extrapolation of the)
theoretical curve.  The typical differences between the observed and
predicted spectral indices are $\sim$0.8, whereas the typical errors in
those EGRET spectral indices are $\sim$0.15; thus the discrepancies are
clearly not due to statistical limitations.  The EGRET-detected FSRQ's
clearly have much harder spectra than the theory predicts.

The RBL's also do not agree well with the theoretical curve, although 
the cluster of points extends both above and below the theoretical curve. 
In this case the discrepancy appears as a broader distribution around
the theoretical curve than would be expected from the errors in the
EGRET indices.  For example, four of the eleven points are more than
2$\sigma$ away from the theoretical curve, whereas statistically no 
more than one would be expected.

It is well-known that the low-energy SED peak frequencies of blazars may
vary with flux levels.  Since most of the SED spectra studies in this paper
were constructed with noncontemporaneous data, one would have to consider 
the possibility
that the lack of correlation between the low-energy SED peak frequency
and the EGRET spectral index in Figure~1 could have been caused by
the shift of the low-energy SED peak positions in different epochs.
We are fairly certain that the EGRET spectra do not change appreciably
with flux (Mukherjee et al.~1997).  As for the shift of the
low-energy SED peak positions, the well-studied sources like those mentioned
at the end of Section 2 indicate that the amount of changes are no bigger
that $\rm \sim$ 1.0 in the scale of $\rm log_{10}(frequency)$, or about
one order of magnitude in frequency.  But if we want to bring the FSRQ or RBL
with hard EGRET spectra in Figure~1 to agree with the theoretical curve,
we would have to shift their low-energy SED peak frequencies by at least
three orders of magnitude.  Such large change in position of the low-energy
SED peak has never been observed, at least for the majority of the blazars
under study in the literature.

Another possible inconsistency between the theoretical models and the 
EGRET data may exist when the spectral shapes of all EGRET blazars 
are viewed together, not restricted to the 27 sources listed in Table~1.
The currently studied theoretical blazar models require that some of the 
EGRET spectra should show clear spectral breaks when the high-energy 
SED peaks pass through the EGRET energy range, much like the spectral
breaks of the five sources detected by all of OSSE, COMPTEL, and EGRET, 
with spectral breaks in SED clearly seen between the OSSE/COMPTEL data 
and the EGRET data (McNaron-Brown et al.~1995).  But in all of the 
spectral fits that have been carried out by the EGRET Team
(Fichtel et al.~1994, Thompson et al.~1995, Thompson et al.~1996,  
Hartman et al.~1999, and references therein), it has never been found
necessary to introduce spectral breaks or spectral cutoffs in order to
properly fit the EGRET data.  It is of course entirely possible that some 
of the EGRET blazar spectra will eventually prove to be more complicated 
than single power laws when the measurements become sufficiently accurate.
In fact, even with the existing EGRET data, Reimer et al.~(1999) and
Bertsch et al.~(1999) are currently carrying out studies to see if 
functional forms more complicated than single power laws will provide
better fit for some of the EGRET blazars.  But we must point out that
in the EGRET energy range, spectral breaks as conspicuous as those
displayed in McNaron-Brown et al.~(1995) or spectral cutoffs well below 
the GeV energy range can be easily recognized in the EGRET data.  
Thus if a theoretical model requires that some of the EGRET sources 
should possess spectral breaks or spectral cutoffs beyond what the 
statistical uncertainties in the EGRET data can accomodate,
then the theoretical model will be inconsistent with EGRET data.

\acknowledgments

     The EGRET team gratefully acknowledges support from the
following: Bundesministerium f\"ur Forschung und Technologie,
Grant 50 QV 9095 (MPE authors); NASA Grant NAG5$-$1742 (HSC);
NASA Grant NAG5$-$1605 (SU); and NASA Contract NAS5$-$31210
(GAC).

\vfil
\eject

\begin{deluxetable}{llllcc}
\tablewidth{0pc}
\tablecaption{EGRET-Detected Blazars Examined in This Study.
\label{table_1}}
\tablehead{
\colhead{Object}      & \colhead{Other} &
\colhead{EGRET Flux} & \colhead{EGRET} &
\colhead{Low-E Peak} & \colhead{Remark} \\
\colhead{}          & \colhead{Name} &
\colhead{E $>$ 100 MeV}      & \colhead{Spectral} &
\colhead{log$_{10}$($f_{peak}$)}      & \colhead{} \\
\colhead{}          & \colhead{} &
\colhead{$\rm 10^{-7}\ cm^{-2}\ s^{-1}$}      & \colhead{Index $\gamma$} &
\colhead{$f_{peak}$ in Hz}      & \colhead{}}

\startdata
0208$-$512 & & $\rm 8.55\pm 0.45$ & $\rm 1.99\pm 0.05$\ \  & 12.0$^{(1)}$ 
 & FSRQ \nl
0219$+$428 & 3C 66A & $\rm 1.87\pm 0.29$ & $\rm 2.01\pm 0.14$ & 13.0$^{(2)}$ 
 & RBL \nl
0235$+$164 & & $\rm 6.51\pm 0.88$ & $\rm 1.85\pm 0.12$ & 13.0$^{(1)}$ & RBL \nl
0420$-$014 & & $\rm 5.02\pm 1.04$ & $\rm 2.44\pm 0.19$ & 12.3$^{(1)}$ & FSRQ \nl
0454$-$234 & PKS & $\rm 1.47\pm 0.42$ & $\rm 3.14\pm 0.47$ & 13.5$^{(2)}$ 
 & FSRQ \nl
0458$-$020 & & $\rm 1.12\pm 0.23$ & $\rm 2.45\pm 0.27$ & 13.0$^{(2)}$ & FSRQ \nl
0521$-$365 & & $\rm 3.19\pm 0.72$ & $\rm 2.63\pm 0.42$ & 13.5$^{(2)}$ & RBL \nl
0528$+$134 & PKS & $\rm 9.35\pm 0.36$ & $\rm 2.46\pm 0.04$ & 12.0$^{(1)}$ 
 & FSRQ \nl
 & & & & & OSSE/COMPTEL \nl
0537$-$441 & & $\rm 2.53\pm 0.31$ & $\rm 2.41\pm 0.12$ & 13.5$^{(1)}$ & RBL \nl
0716$+$714 & & $\rm 1.78\pm 0.20$ & $\rm 2.19\pm 0.11$ & 13.8$^{(3)}$ & RBL \nl
0735$+$178 & & $\rm 1.64\pm 0.33$ & $\rm 2.60\pm 0.28$ & 13.7$^{(2)}$ & RBL \nl
0829$+$046 & & $\rm 1.68\pm 0.51$ & $\rm 2.47\pm 0.40$ & 13.6$^{(2)}$ & RBL \nl
0851$+$202 & OJ +287 & $\rm 1.06\pm 0.30$ & $\rm 2.03\pm 0.35$ & 13.0$^{(2)}$ 
 & RBL \nl
0954$+$658 & & $\rm 1.54\pm 0.30$ & $\rm 2.08\pm 0.24$ & 13.6$^{(2)}$ & RBL \nl
1101$+$384 & Mrk 421 & $\rm 1.39\pm 0.18$ & $\rm 1.57\pm 0.15$ & 16.0$^{(4)}$ 
 & XBL \nl
 & & & & & TeV Detection \nl
1156$+$295 & 4C +29.45 & $\rm 5.09\pm 1.19$ & $\rm 1.98\pm 0.22$ & 13.0$^{(2)}$ 
 & FSRQ \nl
1219$+$285 & W Comae & $\rm 1.15\pm 0.18$ & $\rm 1.73\pm 0.18$ & 13.5$^{(1)}$ 
 & RBL \nl
1226$+$023 & 3C 273 & $\rm 1.54\pm 0.18$ & $\rm 2.58\pm 0.09$ & 13.5$^{(1)}$ 
 & FSRQ \nl
 & & & & & OSSE/COMPTEL \nl
\tablebreak
1253$-$055 & 3C 279 & $\rm 17.97\pm 0.67$ & $\rm 1.96\pm 0.04$ & 12.6$^{(5)}$ 
 & FSRQ \nl
 & & & & & OSSE/COMPTEL \nl
1510$-$089 & & $\rm 1.80\pm 0.38$ & $\rm 2.47\pm 0.21$ & 13.0$^{(2)}$ & FSRQ \nl
1633$+$382 & 4C +38.41 & $\rm 10.75\pm 0.96$ & $\rm 2.15\pm 0.09$ & 12.3$^{(1)}$
 & FSRQ \nl
1652$+$398 & Mrk 501 & $\rm 3.20\pm 1.30^{(6)}$ & $\rm 1.60\pm 0.50^{(6)}$ & 
 16.7$^{(6)}$ & XBL \nl
 & & & & & TeV Detection \nl
2005$-$489 & PKS & $\rm 1.31\pm 0.46^{(7)}$ & $\rm 1.52\pm 0.24^{(7)}$ 
 & 16.4$^{(8)}$ & XBL \nl
 & & & & & TeV Upper Limit \nl
2155$-$304 & PKS & $\rm 3.04\pm 0.77$ & $\rm 1.71\pm 0.24^{(9)}$ 
 & 17.0$^{(10)}$ & XBL \nl
 & & & & & TeV Detection \nl
2200$+$420 & BL Lacertae & $\rm 3.99\pm 1.16$ & $\rm 2.60\pm 0.28$ &  
 14.0$^{(11)}$ & RBL \nl
2230$+$114 & CTA 102 & $\rm 1.92\pm 0.28$ & $\rm 2.45\pm 0.14$ & 12.1$^{(2)}$ 
 & FSRQ \nl
 & & & & & OSSE/COMPTEL \nl
2251$+$158 & 3C 454.3 & $\rm 5.37\pm 0.40$ & $\rm 2.21\pm 0.06$ & 12.8$^{(1)}$ 
 & FSRQ \nl
 & & & & & OSSE/COMPTEL \nl
\tablerefs{(1) von Montigny et al.~1995; (2) Impey \& Neugebauer 1988;
(3) Ghisellini et al.~1997; (4) Macomb et al.~1995; (5) Wehrle et al.~1998;
(6) Kataoka et al.~1999; (7) Lin et al.~1997; (8) Sambruna et al.~1995; 
(9) Vestrand et al.~1995; (10) Chadwick et al.~1999;
(11) Catanese et al.~1997}
\enddata
\end{deluxetable}

\vfil
\eject

\vfil
\eject

\figcaption[]{EGRET photon spectral index $\rm \gamma - 2$ vs. 
         $\rm log_{10}($low-energy peak frequency in SED)
         for EGRET-detected blazars.  \label{fig_1}}

\vfil
\eject

\begin{figure}
\plotone{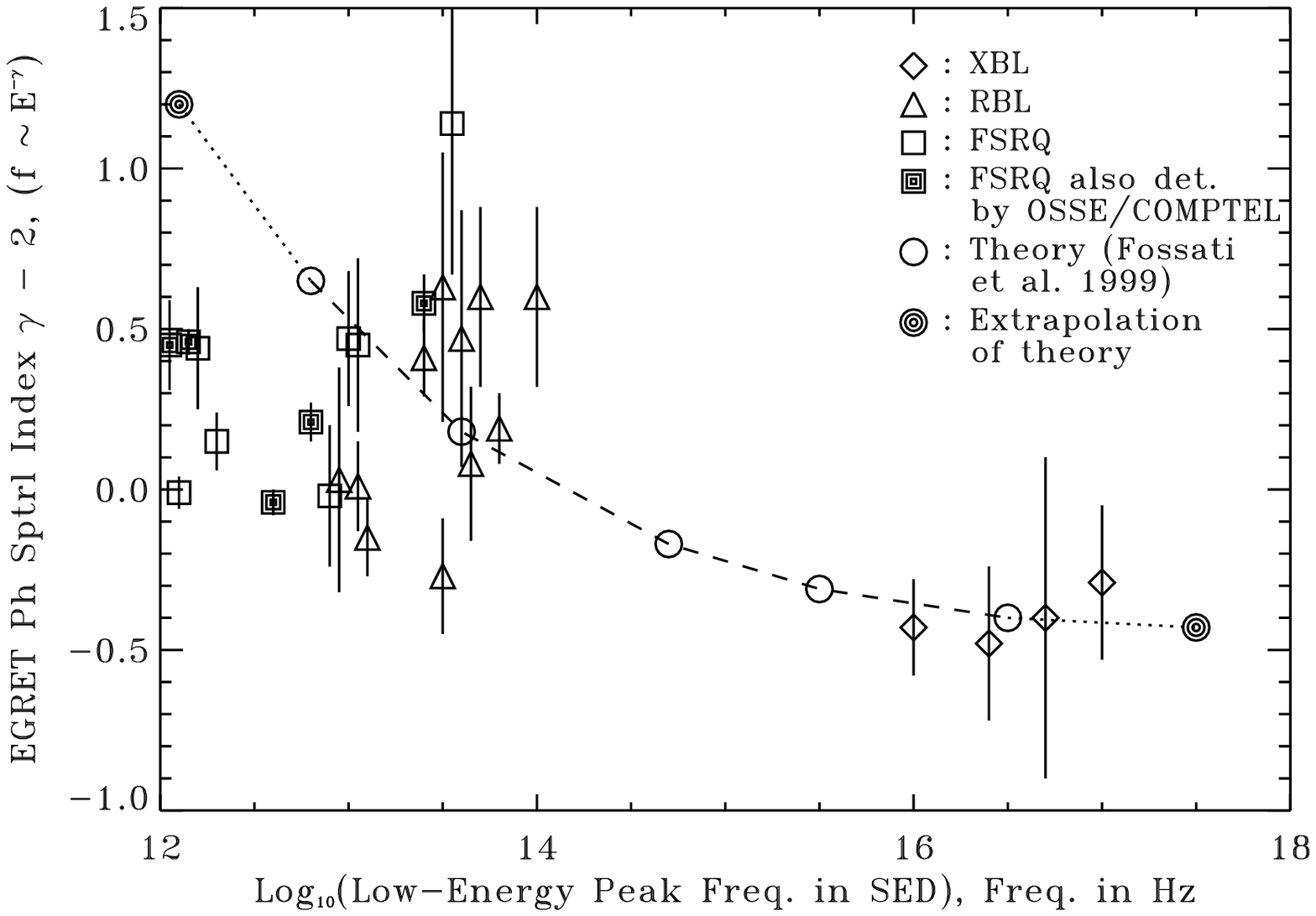}
\end{figure}

\end{document}